\documentstyle[prl,aps,epsfig,amstext,multicol]{revtex}
\begin{document}
\def\am{{$a_{\mu}~$}}
\def\amp{{$a_{\mu ^+}\,$}}
\def\amm{{$a_{\mu ^-}\,$}}
\def\omegaa{{$\omega_a\,$}}
\def\omegap{{$\widetilde{\omega}_p\,$}}

\title{Precise measurement of the positive muon anomalous magnetic moment}
\author{
H.N.~Brown$^2$, G.~Bunce$^2$,
R.M.~Carey$^1$, P.~Cushman$^{9}$,
G.T.~Danby$^2$, P.T.~Debevec$^7$, M.~Deile$^{11}$, H.~Deng$^{11}$,
W.~Deninger$^7$, S.K.~Dhawan$^{11}$, V.P.~Druzhinin$^3$, L.~Duong$^{9}$, 
E.~Efstathiadis$^1$,
F.J.M.~Farley$^{11}$, G.V.~Fedotovich$^3$,  
S.~Giron$^{9}$, F.~Gray$^7$, D.~Grigoriev$^3$, M.~Grosse-Perdekamp$^{11}$,
A.~Grossmann$^6$, 
M.F.~Hare$^1$, D.W.~Hertzog$^7$, V.W.~Hughes$^{11}$,
M.~Iwasaki$^{10}$,
K.~Jungmann$^6$, D.~Kawall$^{11}$, M.~Kawamura$^{10}$, B.I.~Khazin$^3$,
J.~Kindem$^{9}$, F.~Krienen$^1$, I.~Kronkvist$^{9}$,
R.~Larsen$^2$, Y.Y.~Lee$^2$, I.~Logashenko$^{1,3}$,
R.~McNabb$^{9}$, W.~Meng$^2$, J.~Mi$^2$, J.P.~Miller$^1$, W.M.~Morse$^2$,
D.~Nikas$^2$,
C.J.G.~Onderwater$^7$, Y.~Orlov$^4$, C.S.~\"{O}zben$^2$, 
J.M.~Paley$^1$, C.~Polly$^7$, J.~Pretz$^{11}$, R.~Prigl$^2$, 
G.~zu~Putlitz$^6$,
S.I.~Redin$^{11}$, O.~Rind$^1$, B.L.~Roberts$^1$, N.~Ryskulov$^3$,
S.~Sedykh$^7$, Y.K.~Semertzidis$^2$, Yu.M.~Shatunov$^3$,
E.P.~Sichtermann$^{11}$, E.~Solodov$^3$, M.~Sossong$^7$, A.~Steinmetz$^{11}$,
L.R.~Sulak$^1$,
C.~Timmermans$^{9}$, A.~Trofimov$^1$,
D.~Urner$^7$,
P.~von~Walter$^6$, D.~Warburton$^2$, D.~Winn$^5$,
A.~Yamamoto$^8$,
D.~Zimmerman$^{9}$\\
Muon $(g-2)$ Collaboration
}
\address{
$^1$Department of Physics, Boston University, Boston, MA 02215, USA
$~^2$Brookhaven National Laboratory, Upton, NY 11973, USA
$~^3$Budker Institute of Nuclear Physics, Novosibirsk, Russia
$~^4$Newman Laboratory, Cornell University, Ithaca, NY 14853, USA
$~^5$Fairfield University, Fairfield, CT 06430, USA
$~^6$Physikalisches Institut der Universit\"{a}t Heidelberg, 69120 Heidelberg, Germany
$~^7$Department of Physics, University of Illinois at Urbana-Champaign, IL 61801, USA
$~^8$KEK, High Energy Accelerator Research Organization, Tsukuba, Ibaraki 305-0801, Japan
$~^{9}$Department of Physics, University of Minnesota, Minneapolis, MN 55455, USA
$~^{10}$Tokyo Institute of Technology, Tokyo, Japan
$~^{11}$Department of Physics, Yale University, New Haven, CT 06520, USA
}
\date{February 23, 2001}
\maketitle

\begin{abstract}
A precise measurement of the anomalous $g$ value, 
\am$=(g-2)/2$, for the
positive  muon  has  been  made  at  the Brookhaven Alternating Gradient
Synchrotron. The result
$a_{\mu^+}=11~659~202(14)(6)\times10^{-10}~(1.3~\mathrm{ppm})$
is  in good agreement with previous measurements and has an error one
third that of the combined previous data.  
The current theoretical value from the standard model is
$a_{\mu}$(SM)$=11~659~159.6(6.7)\times10^{-10}~(0.57$~ppm) and
$a_{\mu}(\mathrm{exp}) -$$a_{\mu}(\mathrm{SM})=43(16)\times 10^{-10}$ 
in which $a_{\mu}(\mathrm{exp})$ is the world average experimental value. 

\hspace{8.5cm} PACS number: 14.60.Ef ~~~13.40.Em
\end{abstract}

\begin{multicols}{2}
Precise measurement of the anomalous $g$ value, \parbox{2.1cm}{\am$=(g-2)/2$}, 
of the muon
provides  a sensitive test of the standard model of particle physics and
new information on speculative theories beyond it.
Compared to the electron, the muon $g$ value is more sensitive to 
standard model extensions, typically by a factor 
of \( (m_{\mu}/m_e)^2 \).
In  this  Letter  we  report  a  measurement  of  \am  for  the positive
muon from Brookhaven AGS experiment 821, based on data  collected
in 1999. 

The  principle of the experiment, previous results, and many experimental
details have been given in earlier publications~\cite{1998,1997}.
Briefly, highly polarized $\mu^+$ of $3.09 ~\mathrm{GeV/}{\it c}$ 
from a secondary
beamline are injected through a superconducting inflector\cite{krienen} 
into a storage
ring $14.2~\mathrm{m}$ in diameter with an effective circular aperture
$9~\mathrm{cm}$ in diameter. The superferric storage 
ring\cite{danby} has a homogeneous 
magnetic field of $1.45~\mathrm{T}$, which is measured by an NMR system 
relative to the free proton NMR frequency\cite{prigl,fei}. 
Electrostatic quadrupoles provide vertical focusing.
A pulsed magnetic kicker gives a $10~\mathrm{mrad}$ deflection which
places the muons onto stored orbits. 
The muons start in $50~$ns bunches and debunch with a decay time of
about $20~\mu$s due to their $0.6\%$ momentum spread.
Positrons are detected using 24 lead/scintillating fiber electromagnetic
calorimeters\cite{sedykh} read out by waveform digitizers. The waveform 
digitizer and NMR clocks were phase-locked to the Loran C frequency signal.

The muon spin precesses faster than its momentum rotates by an angular
frequency
\omegaa \,in the magnetic field $\left<B \right>$ weighted over the muon 
distribution in space and time. The quantity \am is 
\begin{equation}
a_{\mu} = \frac{\omega_a}{\frac{e}{m_{\mu}c}\left<B\right>},
\end{equation}
where \omegaa is unaffected by the electrostatic field for muons with 
$\gamma=29.3$. Parity violation in the
decay \(\mu^+ \rightarrow e^+ \bar{\nu_{\mu}} \nu_e\) causes
positrons to be emitted with an angular and energy asymmetry. 
Because of the Lorentz boost, the positron emission angle with respect to the 
muon spin direction in the muon rest frame is strongly coupled to its energy
in the laboratory frame.
The number of decay positrons with energy greater than $E$ is described by
\begin{equation}
	N(t) = N_0(E) e^{-t/(\gamma\tau)}
		\left[ 1 + A(E) \sin( \omega_a t + \phi_a(E) ) \right]
	\label{equation:precessionAndDecay}
\end{equation}
in which the time dilated lifetime $\gamma\tau \approx 64.4~\mathrm{\mu s}$. 
Some $140$ $g-2$ periods of $4.37~\mathrm{\mu s}$ were observed.

Most experimental aspects of the data taking in 1999 were the same as 
in 1998\cite{1998}. However, some improvements were
made.  
Care was taken in tuning  the  AGS  ejection  system  to
minimize background from any extraneous
proton beam extracted during the muon
storage time.  
Scintillating  fiber  detectors  which  could be moved in and out of the
storage region were used to study beam properties.
Scintillation  counters  in front of five calorimeters were used
to measure muon losses from the storage ring.
\twocolumn

The  principal  new feature of the 1999 run is the 20-
fold increase in data
collected. During a data acquisition time of 500
hours, we obtained about $2.9$ billion decay positrons with energies greater
than $1.0~\mathrm{GeV}$.
The  AGS  delivered  $4 \times 10^{13}$ protons of $24~$GeV in six 
bunches, separated by $33~$ms, over its $2.5~$s cycle. This resulted in 
$5 \times 10^4$ stored muons per cycle.

\begin{figure}
\center
\epsfig{file=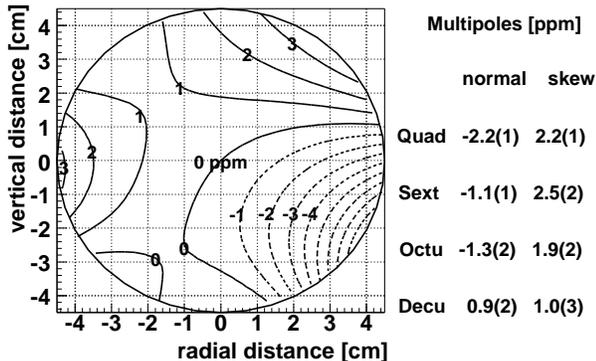, width=0.45\textwidth}
\caption{A 2-dimensional multipole expansion of the field averaged over
azimuth from one out of seventeen trolley measurements.
One ppm contours are shown with respect to a central azimuthal average 
field $B_0=1.451~266~\mathrm{T}$.
The circle indicates the muon beam storage region.
The  multipole  amplitudes at the beam aperture radius of 
$4.5~\mathrm{cm}$ are given. 
}
\label{fig:b_contour}
\end{figure}

The magnetic field $B$ is obtained from NMR measurements of the free proton
resonance frequency $\omega_p$.
Seventeen  NMR  probes are mounted in an array on a trolley which moves
on a fixed track inside  the  muon  storage  ring vacuum.
The  trolley  probes  are  calibrated  with  respect  to  a  standard
spherical H$_2$O probe to an accuracy of $0.2~\mathrm{ppm}$
before and after  data-taking  periods.
Interpolation  of the field in the periods between trolley measurements,
which are made on average every three days, is based  on  the  readings  of
about 150  fixed NMR probes distributed around the ring 
in the top and bottom walls of the vacuum chamber.
Figure~\ref{fig:b_contour}  shows a magnetic field profile averaged over
azimuth. The variations in the amplitudes
of the multipoles affect $\left<B\right>$ by less than $0.02~\mathrm{ppm}$.
The average readings of $36$ uniformly distributed fixed probes are 
maintained to $0.1~\mathrm{ppm}$ by feedback to the main magnet power 
supply.

The $1.45~$T magnetic field of the superconducting inflector is well shielded
from the storage region by flux trapping. However, over an
azimuthal angle of $\sim 1^{\circ}$, a residual fringe field 
reduces  the storage ring central field by about $600~\mathrm{ppm}$,
increasing to $3000~$ppm at the edge of the aperture.
This fringe field reduces the field homogeneity 
at large positive radial and negative vertical distances 
(Fig.~\ref{fig:b_contour}).

The magnetic field is weighted with the muon distribution.
Four thousand simulated muons were tracked for 100 
turns through a field mapped by trolley measurements to 
evaluate the field muons encounter. 
The result agrees with $B$ measured with the trolley probes, 
averaged over the azimuth, and taken at the beam center.

Two largely independent analyses of $\widetilde{\omega}_p$,  
which is proportional to $\left< B \right>$, were made
using different selections of NMR probes.
Their results agree to within $0.03~\mathrm{ppm}$.
The final value is
\(\widetilde{\omega}_p/2\pi = 61~791~256 \pm 25~\mathrm{Hz}~(0.4~\mathrm{ppm})\).
Table  \ref{tb:B_error}  lists the systematic errors.
\parbox{0.48\textwidth}{
\begin{table}
\caption {Systematic errors for the $\widetilde{\omega}_p$ analysis}
\begin{tabular}{l|c} 
Source of errors & Size [ppm] \\
\hline
Absolute calibration of standard probe~~~~~~~~ & $0.05$\\
Calibration of trolley probes &  $0.20$ \\
Trolley measurements of $B_0$ & $0.10$ \\
Interpolation with fixed probes  & $0.15$ \\
Inflector fringe field & $ 0.20$ \\
Uncertainty from muon distribution & $0.12$\\
Others $\dagger$ & $0.15$\\
\hline
Total systematic error on \omegap & $0.4\,~$ \\
\end{tabular}
{\small $\dagger$ higher multipoles, trolley temperature and its power supply 
voltage response, and eddy currents from the kicker.}
\label{tb:B_error}
\end{table}
}

The frequency \omegaa\  is  obtained  from  the  time distribution of
decay $e^+$ counts. The $e^+$ are
detected by calorimeters whose photomultiplier signals 
have a typical FWHM of $5~$ns. 
The signals are sampled every 2.5~ns by 8-bit waveform digitizers (WFD)
with at least $16$ samples per $e^+$ event. The samples are used to 
determine
$e^+$ times and energies, and for pulse overlap studies.
The pulse-reconstruction algorithm fits signals above baseline to an average
pulse shape, determined for each calorimeter individually.
Multiple pulses can be resolved if their separation exceeds 3 to 5~ns.
Systematic  effects  associated  with  the  algorithm  were  extensively
studied.

Whereas  for  the  1998  data, muon decay and spin precession
(Eq.~\ref{equation:precessionAndDecay})
was  adequate  to  describe  the  $e^+$  time
spectrum, the higher count rate and much larger data
sample for 1999 required careful consideration of 
(1) $e^+$ pulses overlapping in time (pileup),
(2) coherent betatron oscillations, 
(3) beam debunching,
(4) muon losses, and 
(5) detector gain stability during the muon storage time.

(1)  The number of pileup pulses in the reconstructed data is proportional
to the instantaneous  counting rate squared and to the minimum pulse separation
time of the reconstruction algorithm.
It amounts to 1\% when the fits of  \omegaa to the data are started, and
distorts  the  $e^+$ time  spectrum  because of misidentification of the
number, energies,  and times  of  the positrons.
Since the phase $\phi_a(E)$ (cf. Eq.~\ref{equation:precessionAndDecay})
depends on the energies of the positrons, pileup has a phase which
differs from $\phi_a$ leading to an error in \omegaa.
Therefore,  the  data are corrected  prior to fitting by subtracting a 
constructed pileup spectrum. 
Positrons found within a window at a fixed time after the pulse that triggers
the WFD are treated as if they overlap with the trigger pulse. The width of
the window is taken equal to the minimum pulse separation time. Only data with
energies at least twice the hardware threshold are fully corrected 
with this method. Our 1~GeV hardware threshold leads to a choice of $E
\ge 2$~GeV in the  \omegaa\ analysis.

The contribution to pileup from signals too small to be reconstructed is not 
accounted for by the procedure described above. These small signals
distort the pulse reconstruction but do not, on average, affect the
energy. However, such unseen pulses introduce small time-dependent
shifts in $\phi_a(E)$ and 
$A(E)$.
The time dependence of the asymmetry, being more sensitive than
the phase, is used to set a limit on the shift of \omegaa.

\begin{figure}
\center
\epsfig{file=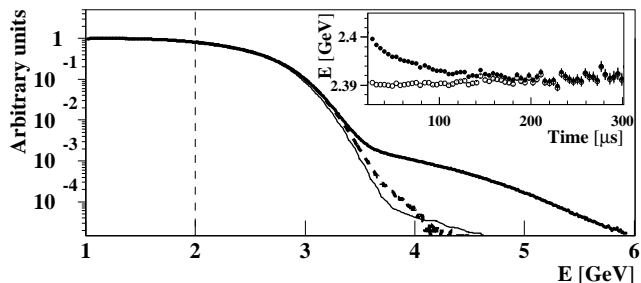}
\caption{
  The energy spectrum of the detected positrons above $1~$GeV
  at all times (thick line) and at only late times (thin line) together
  with the pileup-subtracted spectrum at all times (dashed line). 
  The inset shows the energy above
  $2~$GeV averaged over  one  $g-2$ period as a function of time
  before (filled circles) and
  after (open circles) pileup subtraction for a typical detector.
}
\label{fig:pil}
\end{figure}

Fig.\ \ref{fig:pil} shows the agreement between the positron energy spectrum
after pileup subtraction and the spectrum when pileup is negligible,
together with the uncorrected spectrum.
The inset illustrates that the average energy after pileup subtraction
is constant with time, as expected. 
Without  accounting for pileup,  a shift in \omegaa of $0.3~$ppm would result.

(2) The storage ring is a weak focusing spectrometer
(field index $n=0.137$)
with an aperture which is large compared to the
$18(\rm{w}) \times 57(\rm{h})~\rm{mm}^2$ inflector aperture.
Therefore, the phase space for the betatron oscillations defined by the acceptance
of the storage ring is not filled. In combination with imperfect injection
angles and an imperfect horizontal injection kick, this results in betatron
oscillations of the beam as a whole~ -- ~coherent betatron oscillations (CBO).
These oscillations are observed directly using the fiber monitors. They also
modulate the positron time spectra, since the acceptance of a calorimeter depends
on the muon decay positions. The dominant effect is due to the horizontal 
oscillations, which decay with a time constant of $\sim 100~\mu$s.

(3) The cyclotron period of the bunched beam leads to 
a strong modulation
of the decay positron time spectra which remains at $32~\mu$s when we
begin our fits. The modulation structure is eliminated from the time
spectra by uniformly randomizing the start time for each detector and each
storage fill over the range of one cyclotron period.

(4) Losses  of   muons  during  the  data  collection  are minimized by
controlled scraping\cite{1998} before the data collection is started.
The magnitude and  time-dependence  of  remaining  losses are studied using
coincident  signals  in  the  front  scintillation  counters  of  three
adjacent calorimeters.

(5) Detector gain changes and time shifts are monitored with a pulsed 
laser system and also using the $e^+$ energy spectra. From $32~\mu$s on, 
the gains of the detectors except two
are stable to within 0.1\% over 10 dilated muon lifetimes. The reconstructed 
times are stable to within $20~$ps over 200~$\mu$s (0.1~ppm). 

The raw WFD data were converted into $e^+$ energies and 
times using two independent
implementations of the pulse reconstruction algorithm.
Four independent analyses of \omegaa were made.
For simplicity of presentation one will be described and
the principal differences of the others will be discussed.

\begin{figure}
\center
\epsfig{file=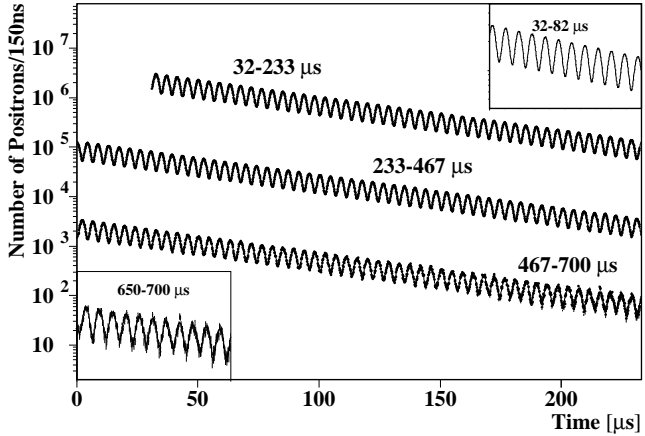}
\caption{Positron time spectrum overlaid with the fitted 10
parameter function ($\chi^2/$dof$=3818/3799$). The total event sample 
of $0.95 \times 10^9~e^+$ with $E \ge 2.0~$GeV is shown.}
\label{fig:wigles}
\end{figure}

The pileup corrected time spectrum for the sum of the detectors is 
fitted by the function,
\begin{equation}
	f(t) = N(t) \cdot b(t) \cdot l(t).
	\label{equation:fitFunction}
\end{equation}
Here, $N(t)$ is the muon decay and precession function of
Eq.~\ref{equation:precessionAndDecay}
and $b(t)$ is the coherent betatron oscillation function,
\begin{equation}
	b(t) = 1 + A_b\, e^{ -t^2 / \tau_b^2 }
		\cos( \omega_b t + \phi_b ),
\end{equation}
and $l(t)$ is the muon loss term,
\begin{equation}
	l(t) = 1 + n_l\, e^{ -t / \tau_l }.
\end{equation}
The quantities $A_b$,   $\tau_b$, $\omega_b$, and $\phi_b$ denote 
the CBO amplitude, time dependence, angular frequency, and phase, respectively,
and $n_l$ and $\tau_l$ denote the fraction of lost muons and its time 
dependence. The CBO frequency $\omega_b$ is 
determined from a Fourier analysis of residuals in a fit of 
Eq.~\ref{equation:precessionAndDecay}
to the data. The remaining 10 parameters in Eq.~\ref{equation:fitFunction} 
are adjusted, in the sense of minimizing $\chi^2$. The frequency \omegaa
correlates strongly only to $\phi_a$. Consequently, \omegaa is insensitive, 
unlike $\chi^2$, to the values of the other 8 parameters and 
to the functional forms of $b(t)$ and $l(t)$.
Fig.~\ref{fig:wigles}  demonstrates  the  good  agreement  of  data and fit, as
indicated by $\chi^2 = 3818$ for 3799 degrees of freedom (dof).

\begin{figure}
\center
\epsfig{file=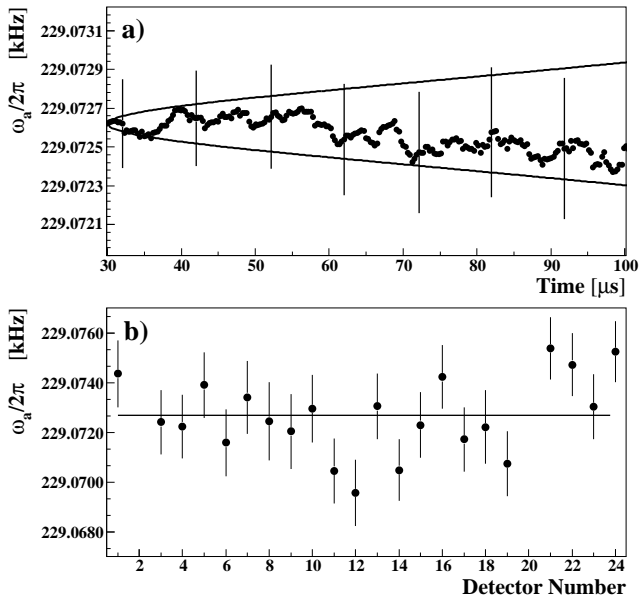}
\caption{a) The fitted frequency $\omega_a/2\pi$ vs. start time of fit 
is shown together with representative error bars. The band indicates the 
size of expected statistical fluctuations. b) Fits to 
the spectra from individual detectors. Detectors 2 and 20 were excluded from
the \omegaa analyses because of a readout problem and a miscalibration, 
respectively.}
\label{fig:consistency}
\end{figure}

The internal consistency of the results was verified in several ways.  
As an example, Fig.~\ref{fig:consistency}a shows the
results when the fit range starts at increasing  times
after  injection and \omegaa is seen to be constant within statistical errors.
Fig.~\ref{fig:consistency}b shows the results for fits to the
spectra from individual detectors ($\chi^2/$dof$=30/21$). The result for 
\omegaa obtained from the average of individual detector fits
(Fig. \ref{fig:consistency}b) is consistent
with the fit to the sum (Fig. \ref{fig:consistency}a) to within $0.07$~ppm.
The  fitted lifetime, after correcting for muon losses, agrees with that
expected from  special  relativity  to  better  than  $0.1\%$.

Two of the three  other   analyses   used  fitting functions similar to
Eq.~\ref{equation:fitFunction}.
The  principal  differences between the analysis described and the others 
consist of a somewhat
different choice of data selection and  fitting  parameters,  a  refined
treatment  of  detector  gain changes, and alternative ways to determine
pileup.  In one analysis, a pileup correction was  made  by  varying  the
minimum  pulse  separation time in the reconstruction algorithm, whereas
in the other analysis pileup is incorporated in the fitted function.
The pileup phase was fixed to the value obtained from the constructed pileup
spectrum. In the fourth analysis, the data are randomly split in four samples
$n_1~$--$~n_4$ which are rejoined in \(u(t)=n_1(t) + n_2(t)\),
\(v(t)=n_3(t-\tau_a/2)+n_4(t+\tau_a/2)\),
and the ratio
\begin{equation}
  r(t) =  \frac{ u(t) - v(t) }{ u(t) + v(t) } = A(E) \sin (\omega_a t+ \phi_a(E)) + \epsilon
\end{equation}
where $\tau_a$ is an estimate of the $g-2$ period and the constant 
$\epsilon \ll 1$.  
This  ratio  is  largely  independent  of  changes of observed counts on
time scales larger than $\tau_a$,  {\it e.g.} the muon lifetime.
The ratio can thus be fitted with fewer free parameters and its results
have somewhat different systematic uncertainties.

The  results  from the analyses are found to agree, on \omegaa\ to within
$0.3~\mathrm~$ppm.
This is within the statistical variation of $0.4~$ppm expected from the
use of slightly different data in the respective analyses.
We combined the results to
$\omega_a / 2\pi = 229072.8 \pm 0.3~\rm{Hz}$~(1.3~ppm).
The only correction applied to our result was $+0.81 \pm 0.08~\rm{ppm}$ for the
effects of the electric field and vertical betatron oscillations\cite{1998}.
The  error  reflects  the  total uncertainty, accounts for the strong
correlations between the individual results, and is  dominated  by  
the statistical contribution.
The systematic errors are listed in Table \ref{tb:omegaa_error}.
The uncorrelated errors were added in quadrature, while the correlated errors 
were added linearly. Most of the systematics were common to all four analyses.
Spin resonances, fit start time, and clock synchronization 
were considered and each was estimated to be less than $0.01~\mathrm{ppm}$.
\parbox{0.48\textwidth}{
\begin{table}
\caption {Systematic errors for the $\omega_a$ analysis.}
\begin{tabular}{l|c} 
Source of errors & Size [ppm] \\
\hline
Pileup & $0.13$\\
AGS background  & $0.10$\\
Lost muons & $0.10$\\
Timing shifts & $0.10$\\
E field and vertical betatron oscillation ~~~~~~~~~~&  $0.08$ \\
Binning and fitting procedure & $0.07$\\
Coherent betatron oscillation & $0.05$\\
Beam debunching/randomization & $0.04$\\
Gain changes & $0.02$\\
\hline
Total systematic error on \omegaa & $0.3$\\
\end{tabular}
\label{tb:omegaa_error}
\end{table}
}

After the \omegap\ and \omegaa\ analyses were finalized, separately and
independently, only then was the anomalous magnetic moment evaluated.
The result is
\begin{equation}
a_{\mu^+}=\frac{R}{\lambda - R}=11~659~202(14)(6) \times 10^{-10}~(1.3~\rm{ppm})
\end{equation}
in which \(R=\omega_a/\widetilde{\omega}_p\), 
\(\mu_{\mu}=e\hbar (1+a_{\mu})/(2m_{\mu}c)\), and
\(\lambda=\mu_{\mu}/\mu_p =3.183~345~39(10)\)\cite{liu}.
This    new    result    is    in    good    agreement   with   previous
measurements\cite{1998,1997,cern} and reduces the  combined  error
by a factor of about three.

The  theoretical  value  of \am in the standard model (SM) \cite{czarnecki} has its dominant contribution from quantum electrodynamics but the weak and 
strong interactions contribute as well. The value
\begin{equation}
a_{\mu}(\mathrm{SM})=11~659~159.6(6.7) \times 10^{-10}\ (0.57~\mathrm{ppm})
\end{equation}
is the sum of  
$a_{\mu}(\rm{QED}) = 11~658~470.56(0.29)\times10^{-10}$
$(0.025~\mathrm{ppm})$,
$a_{\mu}(\mathrm{weak}) = 15.1(0.4) \times10^{-10}$ $(0.03~\mathrm{ppm})$, and
$a_{\mu}(\mathrm{had}) = 673.9(6.7) \times10^{-10}$ $(0.57~\mathrm{ppm})$.
The term $a_{\mu}(\rm{QED})$ is obtained using 
the value of $\alpha$ from \(a_e(\exp)=a_e(\mathrm{SM})\)
\cite{czarnecki},
and terms through order $\alpha^5$ are included. 
The term $a_{\mu}$(weak) includes electroweak contributions 
of up to two-loop order. The term 
$a_{\mu}(\mathrm{had})$ arises from virtual hadronic contributions to the
photon propagator in $4^{th}$ order and $6^{th}$ order, 
where the latter includes 
light-by-light scattering. The $4^{th}$ order term, which provides the largest
contribution and uncertainty to $a_{\mu}(\mathrm{had})$ is obtained from
measured hadron production cross sections in $e^+e^-$ collisions and 
hadronic $\tau$ decay\cite{davier}.
Additional data on $e^+e^-$ collisions 
from Novosibirsk\cite{akhmetshin} 
and from Beijing\cite{zhao} and on $\tau$ decay from Cornell\cite{anderson}
have not yet been
included in the evaluation of $a_{\mu}(\mathrm{had})$.

\begin{figure}
\center
\epsfig{file=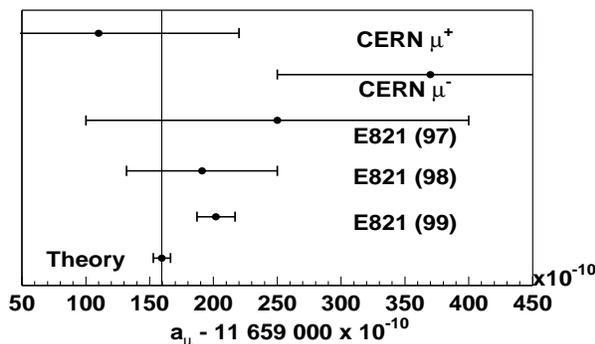, height=5cm, width=8.5cm}
\caption{Measurements of $a_\mu$ and the standard model prediction 
with their total errors.}
\label{fig:a_mu}
\end{figure}

In Fig. \ref{fig:a_mu}, the five most recent measurements of \am are  shown
along with the standard model prediction.  
The difference between the weighted mean of the experimental results, 
$a_{\mu}(\mathrm{exp})=11~659~203(15)\times  10^{-10}~(1.3~\mathrm{ppm})$, and
the theoretical value from the standard model is
\begin{equation}
a_{\mu}(\exp) - a_{\mu} (\mathrm{SM})=43(16)\times 10^{-10}.
\end{equation}
The error is the addition in quadrature of experimental
and theoretical uncertainties.  The difference  is  $2.6$ times the stated
error. 

Many  speculative  theories  predict  deviations from the standard model
value for \am\cite{czarnecki}.
These  include  supersymmetry,  muon  substructure,  and  anomalous   $W$
couplings. The muon anomalous $g$ value is particularly sensitive to 
supersymmetry\cite{moroi} whose contributions to \am come 
from smuon-neutralino and sneutrino-chargino loops. 
In the limit of large tan$\beta$, which is the ratio of the vacuum
expectation values of two Higgs doublets, 
and for a degenerate spectrum of superparticles with mass $\widetilde{m}$,
\begin{equation}
  a_{\mu}(\mathrm{SUSY}) \approx 140 \times 10^{-11}
                         \left(\frac{100~\mathrm{GeV}}{\widetilde{m}}\right)^2
                         \tan\beta .
\end{equation}
If we ascribe the difference $a_{\mu} (\exp) - a_{\mu}(\mathrm{SM})$ 
to $a_{\mu}(\mathrm{SUSY})$, for $\tan\beta$ in the range 4 -- 40, 
then $\widetilde{m} \approx 120$ -- $400$~GeV.

In 2000, approximately four times the total number of positrons were recorded as
compared to our 1999 data. Measurements are now underway
with  negative  muons, which will provide a sensitive test of CPT violation.

We  thank T.~Kirk, D.I.~Lowenstein, P.~Pile, and the  staff of the BNL AGS
for the strong support they have given this experiment. We thank D.~Cassel,
A.~Czarnecki, M.~Davier, T.~Kinoshita, W.~Marciano, and J.~Urheim for
helpful discussions. 
This  work  was supported in part by the U.S.  Department of Energy, the
U.S.  National Science Foundation,  the  German  Bundesminister  f\"{u}r
Bildung und Forschung, the Russian Ministry of Science, and the US-Japan
Agreement in High Energy Physics.

\end{multicols}
\end{document}